\documentclass{PoS}
\usepackage[utf8]{inputenc}  
\usepackage{xspace}
\usepackage{enumitem}   

\newcommand*\aap{A\&A}

\newcommand*\apjl{ApJ}

\newcommand*\nat{Nature}

\newcommand\diff{cm$^{-2}$s$^{-1}$TeV$^{-1}$\xspace}

\newcommand\degr{$^{\circ}$\xspace}

\newcommand\tit{\mbox{\emph{Top :}}\xspace}
\newcommand\bit{\mbox{\emph{Bottom :}}\xspace}

\newcommand\hess{H.E.S.S.\xspace}
\newcommand\hj{HESS J1641$-$463\xspace}
\newcommand\rx{RX J1713.7$-$3946\xspace}

\title{Searching for PeVatrons in the CTA Galactic Plane Survey}

\ShortTitle{Searching for PeVatrons in the CTA Galactic Plane Survey}

\author{\speaker{C. Trichard}$^\dag$, for the CTA Consortium\\
        $^\dag$Aix Marseille Université, CNRS/IN2P3 CPPM, Marseille, France \\
        E-mail: \email{trichard@cppm.in2p3.fr}}

\abstract{The Cherenkov Telescope Array (CTA) will perform a survey of the whole Galactic disk with unprecedented sensitivity at energies up to 300 TeV. One of the key science projects of the CTA consortium is the discovery of Galactic PeVatrons (cosmic ray accelerators to PeV energies). The determination of efficient criteria to identify PeVatron candidates during the Galactic plane survey observations is essential in order to trigger deeper observations. This contribution presents a method which relies on the broadband spectrum of the source to investigate high energy spectral features. The application of this method to specific sources will also be presented.}

\FullConference{35th International Cosmic Ray Conference --- ICRC2017\\
		10--20 July, 2017\\
		Bexco, Busan, Korea}

\begin{document}

\section{Introduction}

Recent measurements of incoming cosmic rays (CR) exhibit a remarkably almost continuous spectral power law as a function of energy up to very high energies. A break of spectral index is observed at $\sim 10^{15}$~eV. The origin of this feature, called the ``knee'', is still debated. However, the accelerators of the particles withe energies below the knee are believed to be located inside the Galaxy. In particular, Galactic supernova remnants (SNR) have been considered to be major contributors to Galactic CRs for decades~\cite{origin}. 
The current generation of gamma-ray astronomy instruments have shown that Fermi acceleration mechanisms are taking place in the shocks of several remnants~\cite{2006A&A...449..223A}~\cite{fermi_IC443_2013}. However all observed SNR spectra exhibit either an exponential-like cutoff, or an index break, significantly below 100~TeV~\cite{funk}. This implies that none of these sources can be firmly identified as a PeVatron. Recent measurements towards the Galactic center have shown that gamma-ray diffuse emission is compatible with a steady source accelerating CRs up to PeV energies~\cite{PeVGalCenter}. However, the identification of the putative source remains unclear.

In order to explain the observed CR spectrum to the highest energies, strong evidence for active PeVatrons associated with astrophysical counterparts is required. The expected sensitivity and angular resolution of CTA above tens of TeV is several orders of magnitude better than current experiments. CTA is therefore a perfect tool to detect and study PeVatrons within our Galaxy.

\subsubsection*{CTA}

CTA is the next generation of Imaging Atmospheric Cherenkov Telescope array (IACT). It is composed of two sites in the northern and southern hemispheres, which will allow observations covering the entire Galactic plane and a large fraction of the extragalactic sky. The array includes three telescope sizes to maximize the energy range of the instrument. The southern site is planned to host around 70 small-sized telescopes, of 4~m in diameter. These telescopes have a large field of view ($\sim$8\degr) and are widely spaced to provide a large array area. This sub-array is therefore the major contributor to the sensitivity of the instrument above few TeV, and will provide measurements up to 300 TeV~\cite{proc_gernot}.

\subsubsection*{CTA Galactic plane survey}

CTA will perform a complete survey of the Galactic plane with an unprecedent sensitivity~\cite{proc_galplane}. This program will be composed of two phases, corresponding to two scans of the Galaxy. The pointing strategy will vary with Galactic coordinates, in order to provide deeper exposures on specific regions, such as the Galactic center. On average, the CTA survey will have 2-4 mCrab flux sensitivity (depending on Galactic coordinates) with a minimum at the Galactic center. This program will provide an unparalleled dataset for the PeVatron search.

\section{PeVatron search}


The PeVatron search is a key science program of the CTA consortium. The goal is to address some of the fundamental questions of high energy CR acceleration. In particular, CTA will provide information on the distribution of PeVatrons in the Galaxy, the mechanism and effectiveness of CR acceleration at PeV energies, and determine if PeVatrons are, as currently expected, commonly associated with young SNR. 


The PeVatron program is composed of two parts. One is dedicated to the observation of \rx. The goals of this observations are to understand basic acceleration processes in amplified magnetic field shocks in one hand, and to investigate the surrounding for run-away particles in the other hand. The second (and main) part is focused in deep observations of the best candidates resulting from observations obtained on the Survey program. We foresee an estimated number of 5 of those candidates, for which an extra 50 h will be devoted to study the highest energy features of their spectra.   

To prepare for this study, knowledge of the sensitivity of CTA to high-energy spectral features and its angular resolution performance is essential. The determination of the criteria which will be used to define the best PeVatron candidates, and the time allocated to each of them, strongly depends on CTA performance in the high-energy range. The goal of this work is to estimate the observation duration required to study high-energy spectral features depending on source spectral properties.

 \subsection*{The method}

This work uses the latest CTA instrument response functions (IRFs) to compute PeVatron search sensitivities. The high-energy bound of CTA IRFs is 200 TeV, due to Monte Carlo simulation limitations~\cite{proc_gernot}. The study of the impact of the limitation of the IRFs on the results is on-going. The results presented here are obtained for zenith angle observation of 20\degr on-axis and for the southern site. The northern part of the array will not host any small-sized telescopes. The performance of CTA above 5 TeV is dominated by this telescope design, and so the performance of the PeVatron search in the northern sky will be significantly worse than in the south, although still far better than that allowed by current arrays.

\begin{figure*}[!ht]
 \centering
  \includegraphics[width=0.49\textwidth]{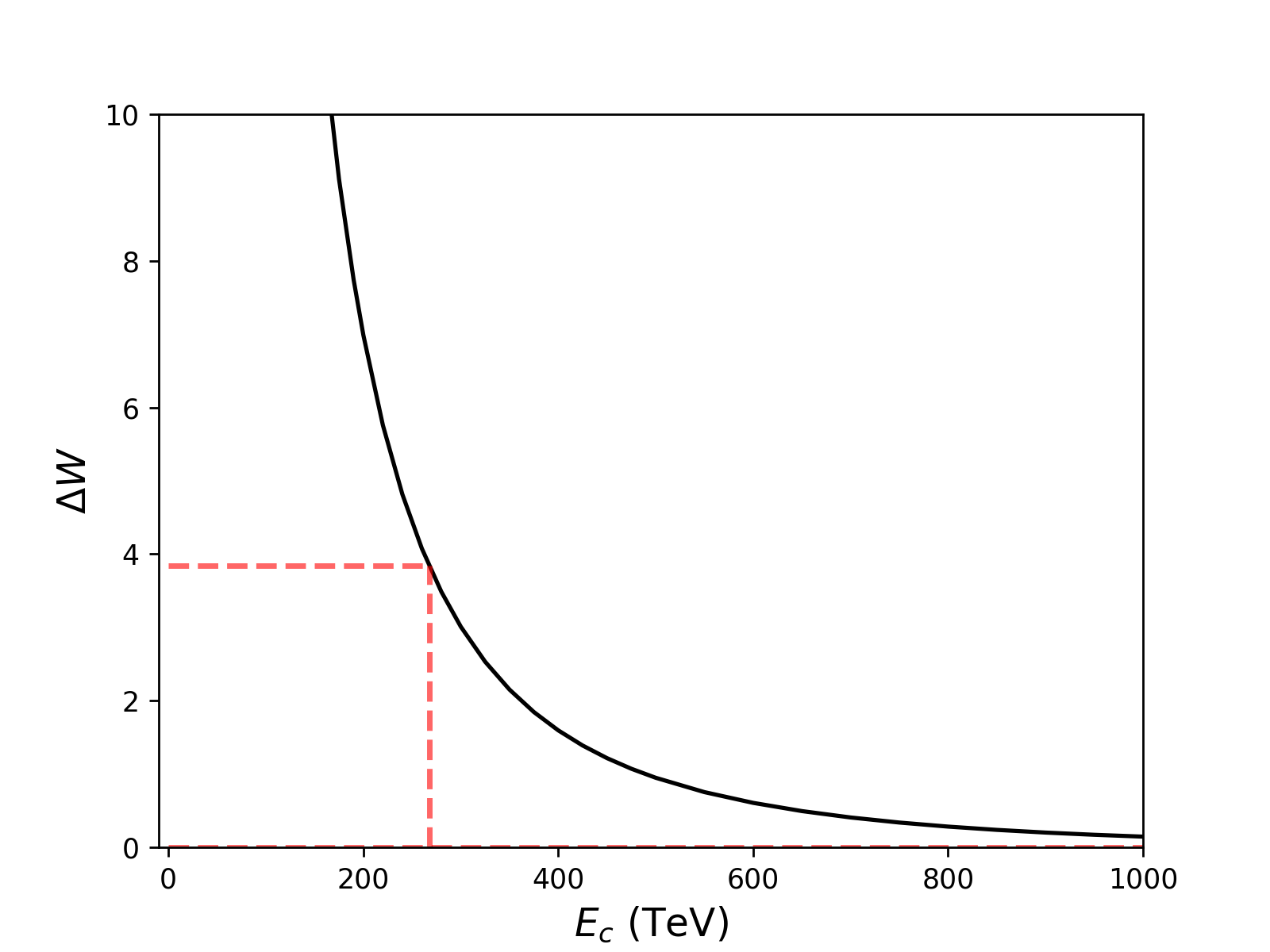} 
      \caption{Profile of W statistics difference, equivalent to a likelihood profile, between the fitted power law model and the same power law with an exponential cutoff, as a function of cutoff energy. This example was computed for a source with $\Phi_0=5.0\,\, .10^{-13}$~\diff, $\Gamma=2$ and 30h of observation.\  \label{plotlike}}
\end{figure*}

The python package Gammapy~\cite{gammapy} version 0.6 was used to :

\begin{enumerate}[label=\textbullet,leftmargin=2cm]
\item Simulate CTA events
\item Fit spectral models on simulated data
\item Derive a goodness-of-fit for each model
\end{enumerate}

To simulate PeVatron sources, simulated events were generated following a power-law spectrum $\Phi(E) = \Phi_0 . \left(\frac{E}{E_0}\right)^{-\Gamma}$, where $\Phi_0$ and $\Gamma$ are the flux normalization and spectral index respectively. The same model was then used to fit the simulated data with the "wstat" statistic from the sherpa fitting module\footnote{http://cxc.harvard.edu/sherpa/statistics/}. To derive a lower limit on the cutoff energy observed by CTA, a model with an exponential cutoff power law (ECPL), $\Phi(E) = \Phi_0 . \left(\frac{E}{E_0}\right)^{-\Gamma} e^{-\left(\frac{E}{E_c}\right)}$, where $E_c$ is the cutoff energy, is compared to the power law fitted model. The cutoff lower limits are determined at 95\% confidence level, i.e. when the difference of the "wstat" statistics, $\Delta W$, is equal to 3.84. The profile of $\Delta W$ is shown on Fig~\ref{plotlike} for one simulation of a source with $\Phi_0=5.0\,\, .10^{-13}$~\diff, $\Gamma=2$ and 30h of observation. The cutoff lower limit is indicated by the red dashed line. 

Note that the distribution of $\Delta W$ does not exactly follow a $\chi^{2}$ distribution with one degree of freedom. The distribution is narrower at lower values. This implies that a limit derived with $\Delta W=3.84$ is at a slightly higher confidence level, of the order of two percent. However, the results presented in the following are derived on the assumption that the test statistic used follow a $\chi^{2}$ distribution with one degree of freedom.

 \subsection*{Results}

\begin{figure*}[!ht]
 \centering
  \includegraphics[width=0.85\textwidth]{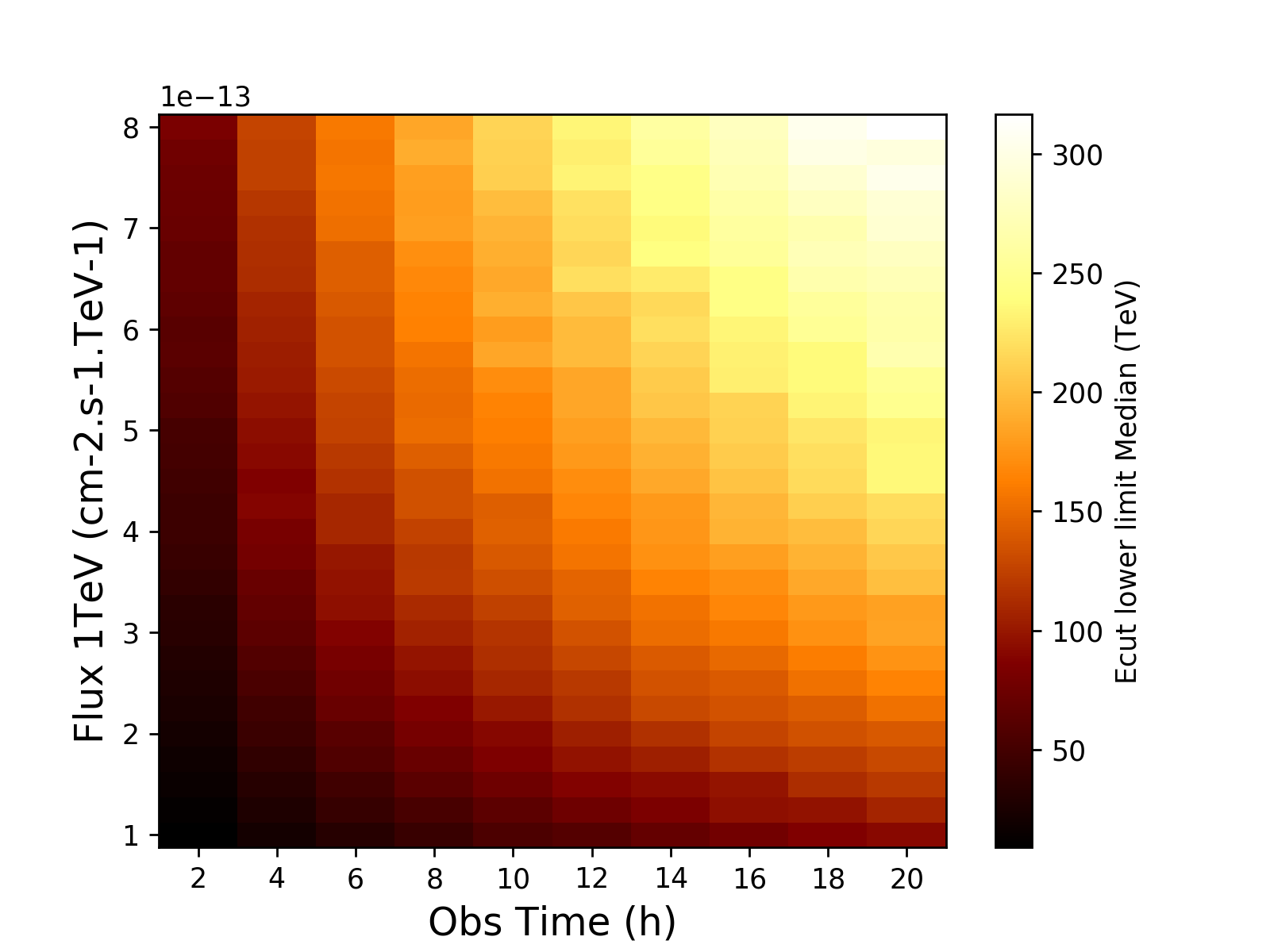} 
  \includegraphics[width=0.85\textwidth]{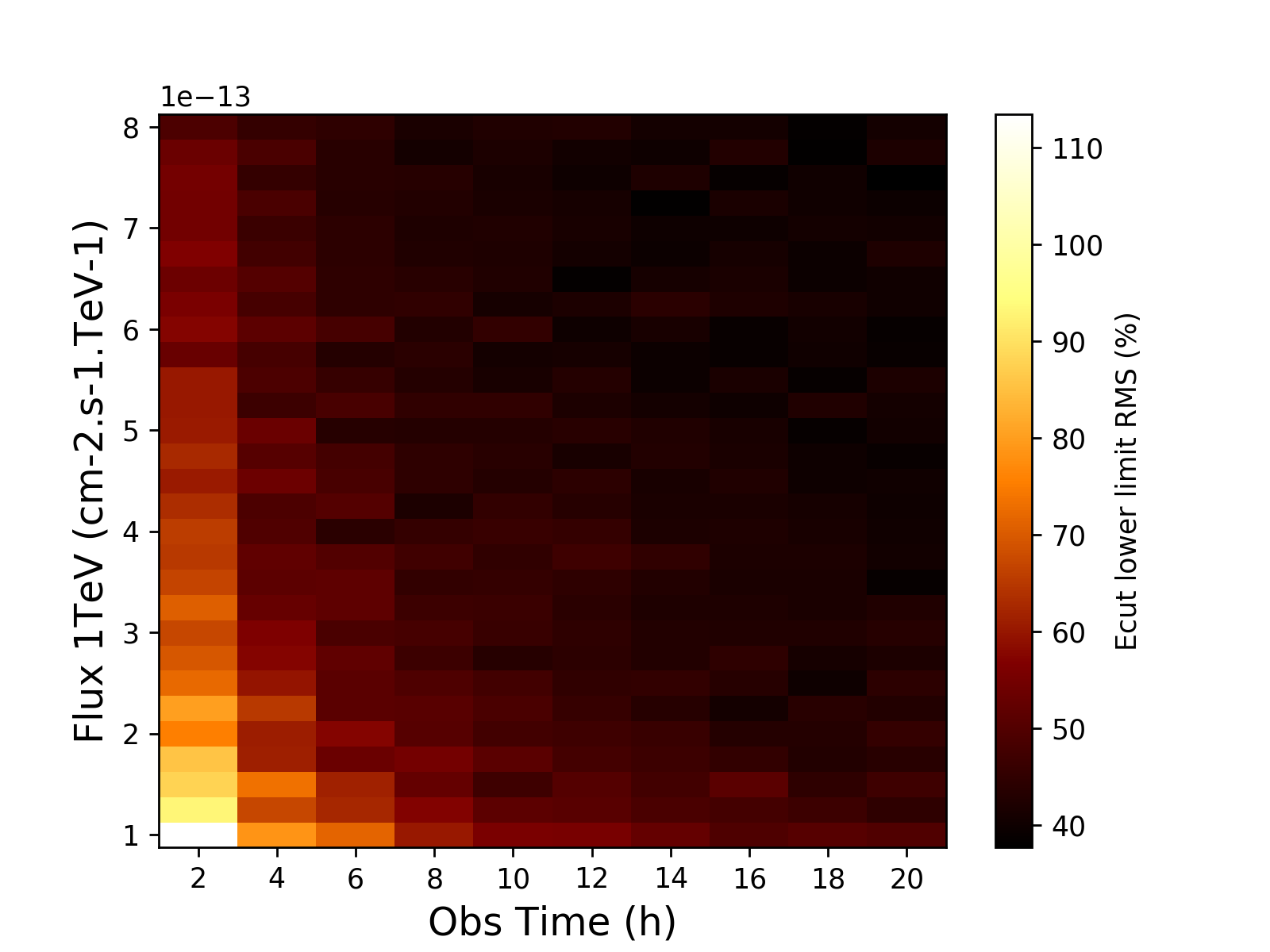} 
      \caption{\tit Median of the cutoff energy lower limit (at 95\% confidence level) distribution in color scale as a function of flux normalization at 1 TeV and observation time.  \bit Same as top, the color scale represents the 68\% width of the distribution. \label{plot1}}
\end{figure*}

The procedure is repeated 1000 times per flux normalization and observation time bin. The outcome distribution spread is due to statistical fluctuations. To illustrate the CTA efficiency for searching for PeVatrons, the median of the distribution is extracted. Figure~\ref{plot1}-(top) shows this median for a source normalization flux from $1$ to $8\,\,.10^{-13}$~\diff and for up to 20 hours of observations. A cutoff lower limit of 100 TeV is obtained with $\sim$8 hours for a source with $\Phi(1TeV)=3.0\,\,.10^{-13}$~\diff and with $\sim$16 hours for a source two times fainter. The Galactic plane survey is expected to have a mean effective exposure of $\sim$15 hours. A lower limit greater than 100 TeV at 95\% C.L. is obtained, with this observation duration, for sources with a flux $\Phi(1TeV)>1.65\,\,.10^{-13}$ \diff.

Figure~\ref{plot1}-(bottom) displays the spread of the distributions of the cutoff lower limits. The width of the distribution for a source with $\Phi(1TeV)=3.0\,\,.10^{-13}$~\diff, for 6 and 20 hours of observation, are $\sim$49\% and $\sim$~44\%, respectively.

\begin{figure*}[!ht]
 \centering
  \includegraphics[width=0.80\textwidth]{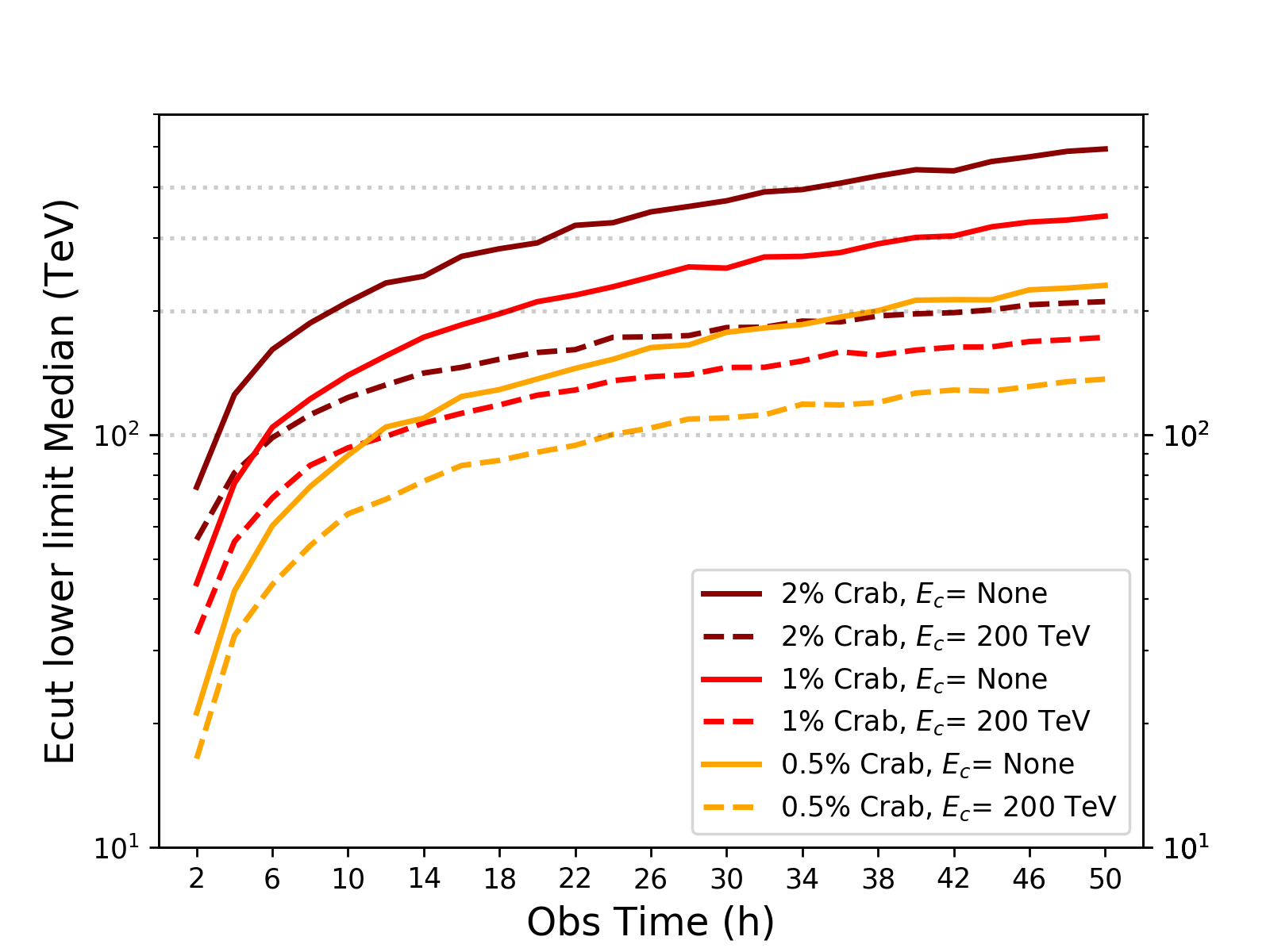} 
      \caption{Median of the cutoff energy lower limit (at 95\% confidence level) distribution as a function of observation time. The continuous and dashed lines represent simulations without cutoff energy and with a cutoff energy at 200 TeV respectively. \label{plot2}}
\end{figure*}

PeVatrons accelerate particles up to PeV energies. However, the existence of a cutoff above a few PeV is expected. This would imply a cutoff of the gamma-ray spectrum at a few hundreds of TeV. Such a cutoff would affect the determination of the high energy features in CTA-observed spectra. In order to probe this impact, the sources were also simulated with a spectral cutoff at 200 TeV. The same procedure, as described above, has been applied. The results obtained for a source with 2\%, 1\% and 0.5\% Crab flux at 1 TeV ($\Phi(1TeV)^{1 Crab}=3.8\,\, 10^{-11}$~\diff~\cite{2014A&A...562L...4H}) are shown in Fig~\ref{plot2}. 

A cutoff at 200 TeV in the source spectra strongly affects the derived limit of the cutoff energy. For instance, for 10h of observation, the lower limit of the cutoff energy for a 2\% (0.5\%) Crab flux source observed is 41\% (22\%) lower than the one observed when the intrinsic spectra is a pure power law. This effect increases with deeper observations to, for instance, 57\% (40\%) at 50h for the same source flux. This effect could provide a powerful tool to probe upper limits on energy cutoffs of the intrinsic spectrum of the source.

\section{\hj}

\hess is a running IACT experiment located in Namibia, composed of four 13~m and one 28~m diameter telescopes. The \hess collaboration produced a survey of the Galactic plane revealing more than 70 very-high-energy sources~\cite{HGPS2013}. One of them, \hj, is among the best PeVatron candidates known~\cite{2014ApJ...794L...1A}. The source exhibits a rather hard spectrum, extending to a few tens of TeV without any sign of curvature. The \hess collaboration have used this result to determine a 99\% C.L. cutoff lower limit of the proton spectrum at 100~TeV.  

The capability of CTA to probe possible spectral features at very high energies was investigated. A source with a flux of $\Phi_0=3.91\,\,.10^{-13}$~\diff and $\Gamma=2.07$ was simulated. The method described above was used to determine the cutoff lower limit expected with CTA. The capability of CTA to detect the cutoff of the source was also investigated. A fit with an ECPL model was performed and compared to the fit with a pure power law. The ECPL model is considered to be significantly better at describing the data if $\Delta W>6.635$, equivalent to a $\sim$3$\sigma$ detection.

\begin{figure*}[!ht]
 \centering
  \includegraphics[width=0.75\textwidth]{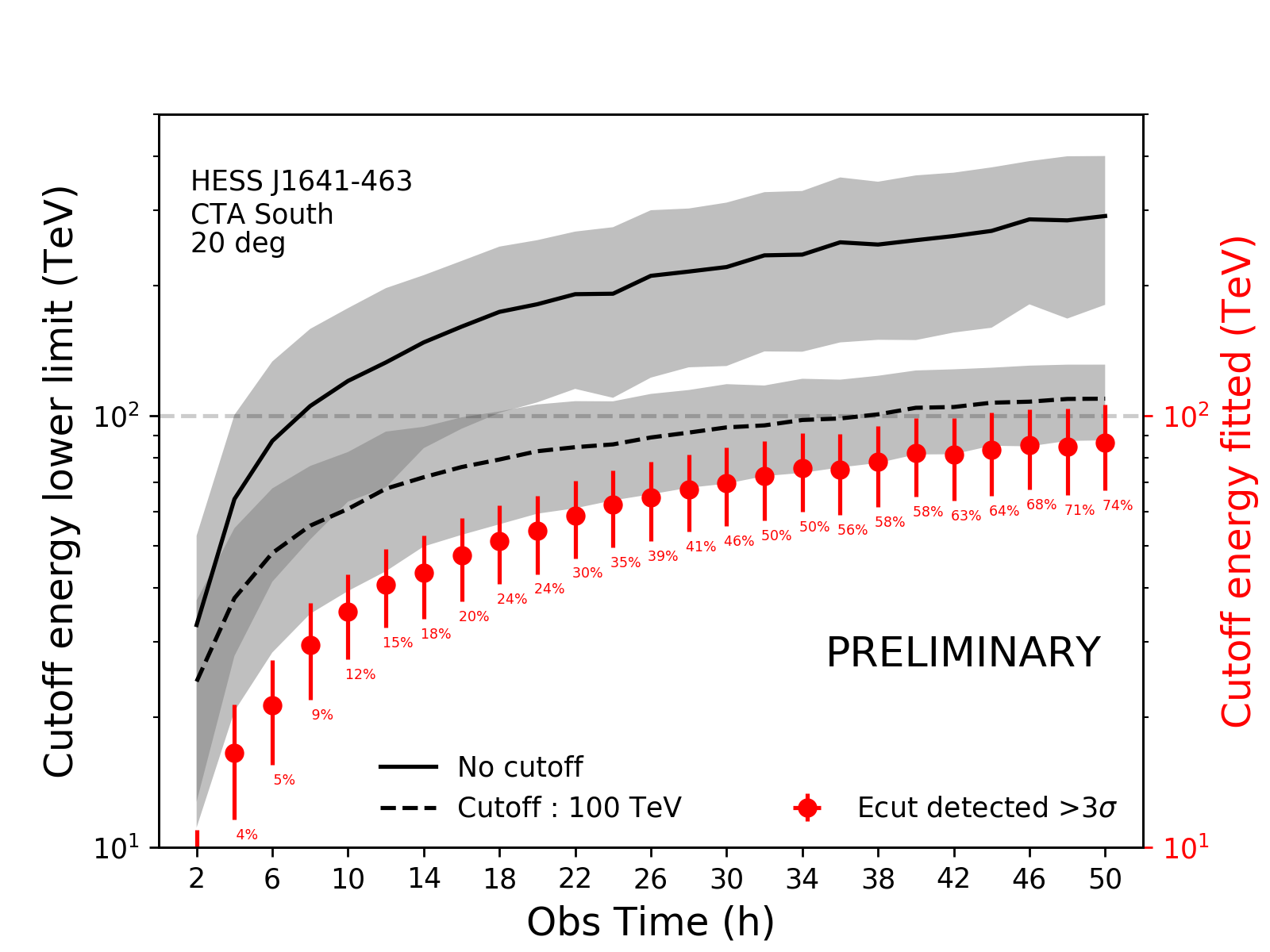} 
      \caption{Median of the cutoff energy lower limit (at 95\% confidence level) distribution as a function of observation time (lines). The continuous and dashed lines represent a source without cutoff energy and with a cutoff energy at 100 TeV (when the fit with an ECPL is less than 3$\sigma$ significant) respectively. The grey bands show 68\% width of the distribution. The red dots are the median of the fitted cutoff energy when the fit with an ECPL is more than 3$\sigma$ significant. The red error bars represent the 68\% width of the distribution. The values indicate the probability that the fit with an ECPL is significant.  \label{plotj1641}}
\end{figure*}

The results are shown in Fig~\ref{plotj1641}. The black lines are the cutoff lower limits for a source simulated with a cutoff at 100~TeV (dashed) and without cutoff (continuous). These lower limits correspond to simulations for which the fit with an exponential cutoff significance is greater than 5$\sigma$. The points represent the distribution of the cutoff energy fitted when the ECPL model fit significance is above 3$\sigma$. The error bars describe the 68\% width of the distribution. The values indicated below the points represent the probability to detect the cutoff above 3$\sigma$ for each observation time.

For this source, a cutoff lower limit of 100 TeV is expected to be reached in $\sim$7h to $\sim$38h depending on the source cutoff energy. The probability of detecting an exponential cutoff after 15h of observation is $\sim$19\% with a median of the cutoff energy distribution of 45~TeV.


\section{Summary}

In summary, the PeVatron search within the Galactic plane is a CTA key science project. This instrument is designed to be well suited to address the fundamental questions about the origin of the highest energy Galactic cosmic rays.  

The study of the performance of CTA final layout for the PeVatron search is on-going. The cutoff lower limit, as a function of source flux, intrinsic spectral features and observation time have been estimated using simulations. A cutoff lower limit of 100~TeV is obtained for a source with $\Phi(1TeV)=3.0\,\,.10^{-13}$~\diff after $\sim$16 hours of observation, which will be approximately the mean Galactic plane exposure.  

For the specific source, \hj, an observation time of $\sim$7h to $\sim$38h is required to put a lower limit of 100~TeV of the cutoff energy.

\section{Acknowledgments}
\noindent
This work was conducted in the context of the CTA Consortium. 

\noindent
We gratefully acknowledge financial support from the agencies and organizations listed here:
http://www.cta-observatory.org/consortium\_acknowledgments, and especially the OCEVU Labex (ANR-11-LABX-0060) and the Excellence Initiative of Aix-Marseille University - A*MIDEX, both part of the French “Investissements d’Avenir” programme. We gratefully acknowledge the LUTH for providing the PowerEdge R730 server.

\end{document}